\documentclass[aps,prb,twocolumn,showpacs,floatfix]{revtex4-2}

\usepackage{graphicx,color}
\usepackage{amsfonts}
\usepackage[figuresright]{rotating}
\usepackage{amssymb}
\usepackage{amsmath}
\usepackage{pifont}
\usepackage{psfrag}
\usepackage{subfigure}
\usepackage{multirow}
\usepackage{tabularx}
\usepackage{textcomp}
\usepackage{units}
\usepackage{hyperref}
\usepackage{comment}
\usepackage{longtable}
\hypersetup{
 pdfnewwindow=true, colorlinks=true,
 linkcolor=blue, anchorcolor=blue,
 citecolor=blue, filecolor=blue,
 menucolor=blue, urlcolor=blue}

\def\nn{\nonumber}

\def\beq{\begin{eqnarray}}
\def\eeq{\end{eqnarray}}

\renewcommand{\v}[1]{\ensuremath{\mathbf{#1}}} 

\let\baraccent=\= 
\renewcommand{\=}[1]{\stackrel{#1}{=}} 
	
\makeatletter

\begin{document}

\title{Bulk photospin effect: Calculation of electric spin susceptibility to second\\ order in an electric field}

\author{Benjamin\ \surname{M. Fregoso}}
\affiliation{Department of Physics, Kent State University, Kent, Ohio 44242, USA}

\begin{abstract}
We compute the electric spin susceptibility of Bloch electrons with spin-orbit coupling to second order. We find that it is possible to generate a nonequilibrium spin polarization in the bulk of non-magnetic inversion-symmetric materials using linearly polarized electric fields, but the process depends on interband coherence and produces heating. It may be possible to avoid heating with circular polarization in certain scenarios. The standard Edelstein effect and spin orientation effects are recovered in appropriate limits within the formalism. Finally, the electric spin susceptibility of metals has contributions proportional to spin multipole moments of the Fermi sea that dominate the low frequency spin response.
\end{abstract}

\maketitle

\section{Introduction and main results}
An electric field can exert torque on an electron's spin through the field of the ions. This spin-orbit coupling (SOC) can be large in materials that break inversion symmetry. The idea of using electric fields to establish a nonequilibrium spin polarization has been extensively explored.~\cite{Roth1960,Meier1984,Bychkov1984,Aronov1989,Edelstein1990,Moessner2021,Zutic2004,Kimel2005,Tarasenko2005,Kato2004,Silov2004,Ganichev2006,Crankshaw2009,Pesin2012,Rioux2012,Raimondi2012,Xu2021} The subsequent evolution of the spin ensemble and the path to reach equilibrium is determined by the balance between spin-injection and spin-relaxation processes. Manipulation of an individual electron's spin or spin ensembles could potentially find applications ranging from information storage to energy-efficient electronics and quantum computation.\cite{Zutic2004,Pesin2012} 

In the presence of a static $\v{E}_{0}$ and an optical $\v{E}$ electric field, we can expand the time-independent (dc) spin polarization in powers of the electric fields as 
\begin{align}
\v{S}_{dc} &=  \zeta_{e}^{(1)} \v{E}_0  + \zeta_{e}^{(2)} \v{E}_0^2 + \zeta^{(2)}_{bpse} \v{E}^2 +\cdots, 
\label{eq:S_Elec_exp}
\end{align}
where $\zeta^{(1)}_{e}$ is the linear dc spin susceptibility, $\zeta^{(2)}_{e}$ is the quadratic dc spin susceptibility, and $\zeta^{(2)}_{bpse}$ is the \textit{bulk photospin effect} (BPSE) susceptibility. In metals, the linear term gives the leading contribution. In insulators, the quadratic term gives the leading contribution. 

Since, under spatial inversion,  the spin magnetization is even but the electric field is odd, the linear spin susceptibility vanishes if there is spatial inversion symmetry (IS). To have a nonzero linear susceptibility, IS must be broken. Indeed, a spin polarization can be established, to linear order, by a static electric field in non-magnetic metals with broken IS, known as the spin Edelstein effect.\cite{Aronov1989,Edelstein1990} This phenomenon can be understood on the basis of the SOC provided by the field of the ions.~\cite{Bychkov1984} The Edelstein effect has been observed experimentally in GaAs\cite{Kato2004,Silov2004,Ganichev2006} and studied in topological insulators,\cite{Pesin2012,Mellnik2014,Ando2014,Li2014,RodriguezVega2017,Acosta2019} van der Waals heterostructures,\cite{Ghiasi2019,Hoque2020}, Weyl semimetals~\cite{Johansson2018,Zhao2020} and superconductors.\cite{He2020,Ikeda2020}

The Edelstein effect is usually associated with metals where the Fermi surface (FS) dominates the spin response at low frequencies. Here we find that the FS contribution to the $m$-th order electric spin susceptibility can be thought, more intuitively, as the average $m$-th spin multipole of the Fermi sea. For example, to first order, the dc spin polarization of metals is proportional to the average spin dipole of the Fermi sea. These FS spin susceptibilities are entirely analogous to higher order Drude conductivities which can be thought as velocity moments of the Fermi sea.

Interestingly, contrary to the linear response, a nonlinear spin polarization can be generated in the bulk of homogenous materials without interfaces. This \textit{bulk photospin effect} (BPSE) is characterized by the quadratic electric spin susceptibility. It is useful to separate the BPSE susceptibility into its symmetric and antisymmetric parts 
\begin{align}
\v{S}_{dc,bpse} = \nu_{2} |\v{E}|^{2} + \upsilon_2 \v{E}\times\v{E}^{*},
\label{eq:Sdc_s_as}
\end{align}
where $\nu_2$ is symmetric in the electric field indices and $\upsilon_2$ is antisymmetric. $|\v{E}|^{2}$ indicates (schematically) a symmetric combination of field indices. The BPSE is analogous to the photovoltaic effect (BPVE)\cite{Sturman1992,Baltz1981,Sipe2000,Tan2016,Ahn2022}, whereby a constant current is generated in materials that lack IS. Since the current is odd under spatial inversion, but the electric field is even, the BPVE vanishes if the point group of the material has IS. To have a nonzero BPVE, the material has to break IS. In the BPSE, on the other hand, both the spin magnetization and the electric field are even under spatial inversion, and hence no restrictions are imposed by IS (to second order). Under the time reversal operation, the spin magnetization and current are both odd, and hence time reversal symmetry (TRS) imposes the same restrictions on both the BPVE and BPSE.

In the BPSE, $\upsilon_2$ characterizes the spin response to fields with circular polarization. $\upsilon_2$ processes have been extensively studied in the context of spin orientation phenomena\cite{Meier1984,Zutic2004,Rioux2012} and are usually associated with angular momentum transfer from circularly polarized light to electrons' spin. $\nu_{2}$, on the other hand, characterizes the generation of spin polarization with linearly polarized light. $\nu_{2}$ processes have received less attention,\cite{Tarasenko2005,Crankshaw2009} perhaps because it is unclear where does the spin angular momentum comes from. Since linear polarized photons do not carry angular momentum, and the material does not break TRS by assumption, an internal torque appears in the system that transfers angular momentum to and from other degrees of freedom, e.g., charge, phonons, excitons, etc. For example, an internal torque could lead to current loops and dissipation. Alternatively, in the absence of an external torque, a spin polarization may be accompanied by a rotation of the sample as a whole (Einstein-de Haas effect). 

In this paper, we present a microscopic derivation of $\nu_2$ and $\upsilon_2$. We consider non-magnetic insulators and metals in any configuration of external field frequencies. For insulators we find that $\nu_2$ depends on the off-diagonal elements of the density matrix and hence requires quantum coherence. $\nu_2$ vanishes when field frequencies are lower than the energy gap (it is resonant), and hence the system absorbs energy and heats up. $\upsilon_2$, on the other hand, has resonant and nonresonant contributions. The former depends on the diagonal elements of the density matrix and gives the usual spin orientation phenomena.\cite{Meier1984,Zutic2004,Rioux2012} The latter means that it is possible to generate spin polarization with circularly polarized light without producing heat.\cite{Kimel2005} 

We solve the Boltzmann equation pertubatively in the electric field with a simple collision integral in the relaxation time approximation. Although specific diagrammatic approaches are effective,\cite{Parker2019} the Boltzmann equation is naturally conserving\cite{Kadanoff1962} and easily interpreted physically. We follow a first principles approach in the sense that the details of the Bloch matrix elements are hidden and work only with relations among Bloch matrix elements. This perspective is suited to finding common features across material applications and gives explicit expressions for response functions that can later be used in large-scale numerical codes. Because only the bare minimum model of dissipation is considered, our approach does not include effects whose origin lies in disorder;\cite{Raimondi2012,Atencia2021} rather, the origin of the phenomena we describe here is in field-matter interactions.

Phenomena related the BPSE have been studied before in specific cases. For example, photomagnetization by circularly polarized light (inverse Faraday effect (IFE)) has been extensively studied theoretically using semi-classical formalisms~\cite{Pitaevskii1961,Zon1976,Hertel2006,Woodford2009}, quantum mechanical formalisms~\cite{Pershan1966} based on model systems~\cite{Kurkin2008,Taguchi2011,Popova2011,Popova2012,Gridnev2013,Qaiumzadeh2013}, a quantum mechanical formalism~\cite{Battiato2014} with applications to real materials~\cite{Berritta2016}, a first-principles calculation by Keldysh formalism~\cite{Freimuth2016,Freimuth2021}, a diagramatic perturbation theory of Rashba model and recently, a first-principles formalism of nonlinear response~\cite{Xu2021}. 

One difference with previous pioneering works is that we do not consider the contribution from the orbital magnetization $-g_L\mu_B \v{L}/\hbar$ because the operator $\v{L}=\v{r}\times \v{p}$ in the Bloch basis is very singular~\cite{Thonhauser2005,Resta2010,Essin2009,Essin2010,Aryasetiawan2019}. In this case a separate approach is preferable~\cite{Shi2007}. In practice, it maybe possible to devise approximate schemes~\cite{Go2020,Go2021}. In addition, we do not consider the $g_s-$factor renormalization of the spin magnetization $-g_s \mu_B \v{S}/\hbar$ expected to occur in a crystal~\cite{Roth1960}. Here we focus on the \textit{spin polarization} not on the magnetization itself. 

A second difference with previous works is that our formalism takes into account intraband and interbad processes on equal footing. This leads to new phenomena. For example, it is possible to induce a spin polarization with linearly polarized light because of quantum interband coherence; in stark contrast to the IFE which requires circularly polarized light. Similarly, we find that metals have FS-specific intraband contributions to the spin polarization which are proportional to spin multipoles of the Fermi sea and which dominate the low frequency response. 

The paper is organized as follows. Sec.~\ref{sec:notation} describes the notation used in this paper. In Sec.~\ref{sec:Boltzmann_eqn}, we solve the Boltzmann equation for the density matrix up to second order in the electric field. These solutions are then used to construct the general first-order susceptibility (Sec.~\ref{sec:zeta1_gen}), apply it to special cases (Sec.~\ref{sec:zeta1_special_cases}), in particular, to a TI in an electric field (Sec.~\ref{sec:zeta1ej}). We then construct the general second-order susceptibility (Sec.~\ref{sec:zeta2_general}), apply it to special cases (Sec.~\ref{sec:zeta2_special_cases}), in particular, the dc quadratic Edelstein susceptibility (Sec.~\ref{sec:zeta2e000}), the BPSE in insulators (Sec.~\ref{sec:bpse_ins}) and metals (Sec.~\ref{sec:bpse_met}). Then we give an example of a spin quadrupole in a TI under a magnetic and an electric field (Sec.~\ref{sec:example_ti_mag_elec}), and of optical spin coherence of conduction bands (Sec.~\ref{sec:opical_inj_ex}). A final discussion is presented in Sec.~\ref{sec:conclusion}.

\section{Notation}
\label{sec:notation}
We follow the notation of Ref.~\cite{Fregoso2019}. Spin response functions are written as $\zeta^{abc...}(-\omega_{\Sigma},\omega_\beta,\omega_{\delta},...)$, where $abc...$ are Cartesian indices, $\omega_\beta,\omega_{\delta},...$ are frequency components of the external electric field, and $\omega_{\Sigma}$ is the sum of those frequencies. Bold fonts represent vectors. The covariant derivative is denoted with a semicolon, e.g., $\v{s}_{nm;b}$ is the covariant derivative of the $nm$ matrix element of the spin in the $b$ Cartesian direction. The Bloch state $|n, i_n,\v{k}\rangle$ is denoted as $|n,\v{k}\rangle$, where the band index $n$ and spinor index $i_n$ are lumped together, and $\v{k}$ is the crystal momentum. The time-reversed state is denoted as $|n,\bar{i}_n,-\v{k}\rangle$ or simply $|\bar{n},-\v{k}\rangle$. $\bar{i}_n$ is the spin-flipped spinor.

When the meaning is clear from the context, we often omit frequency arguments, Cartesian indices, the crystal momentum, and time dependance of operators from response functions. A detailed summary of definitions is given in Appendix~\ref{sec:defs}.

\section{Boltzmann equation}
\label{sec:Boltzmann_eqn}
We consider a classical homogenous electric field with multiple frequency components $\omega_{\beta}$
\begin{align}
E^{b}(t) = \sum_{\beta} E^{b}_{\beta} e^{-i\omega_{\beta}t},
\end{align}
acting on an ensemble of Bloch electrons characterized by the density matrix $\rho_{mn}$. The density matrix evolves according to the Boltzmann equation
\begin{align}
\frac{\partial \rho_{mn}}{\partial t} + i \omega_{mn}\rho_{mn} -\frac{e}{i\hbar}\sum_{lb} E^{b}(\rho_{ml}r^{b}_{ln} - r^{b}_{ml}\rho_{ln}) \nn \\
+ \frac{e}{\hbar}\sum_{b} E^{b}\rho_{mn;b} = -\frac{1}{\tau}(\rho_{mn} - \rho^{(0)}_{mn}).
\label{eq:Boltzmann_eom}
\end{align}
The notation is given in Appendix~\ref{sec:defs}. The left hand side describes coherent motion due to the electric field. It is obtained from the equation of motion of $\hat{\rho}$, see Ref.~\cite{Fregoso2019}. The right hand side is added phenomenologically to describe dissipative processes. It is a collision integral in the relaxation time approximation. Eq.~\ref{eq:Boltzmann_eom} incorporates interband and intraband matrix elements on an equal footing. This guarantees, among other things, Maxwell's equation $d\v{P}/dt = \v{J}$ holds in the Bloch basis ($\v{P}$ electric polarization, $\v{J}$ electric current).\cite{Fregoso2019} Also, Eq.~\ref{eq:Boltzmann_eom} takes into account interband coherence important to recover, e.g., the Hall conductivity,~\cite{Culcer2017} shift current,\cite{Baltz1981,Fregoso2019,Orenstein2021} etc. If only intraband processes are important, $\rho_{nm} \to \delta_{nm} \rho_{nn}$, Eq.~\ref{eq:Boltzmann_eom} reduces (as expected) to the one-band semiclassical Boltzmann equation 
\begin{align}
\frac{\partial \rho_{nn}}{\partial t} + \frac{e}{\hbar} \v{E} \cdot \pmb{\nabla}_{\v{k}} \rho_{nn}  = -\frac{1}{\tau}(\rho_{nn}- \rho^{(0)}_{nn}).
\label{eq:Boltzmann_semicl}
\end{align}
Momentum relaxation produces spin relaxation via SOC mechanism, e.g., Dyakonov-Perel', but the details of such process are not considered here. Our relaxation time does not depend on momentum or energy. We solve Eq.(\ref{eq:Boltzmann_eom}) in powers of the electric field as 
\begin{align}
 \rho_{mn} = \rho_{mn}^{(0)} + \rho_{mn}^{(1)} + \rho_{mn}^{(2)} + \cdots,
\end{align}
where $\rho_{mn}^{(0)}$ is the density matrix in the absence of fields and $\rho_{mn}^{(n)}$ are higher order terms. In the long-time limit we obtain
\begin{align}
\rho_{mn}^{(0)} &= \delta_{nm}f_n, \\
\rho_{mn}^{(1)} &=\sum_{b\beta} \bar{\rho}^{(1)b\beta}_{mn} E^{b}_{\beta}e^{-i\omega_{\beta}t}, \\
\rho_{mn}^{(2)} &=\sum_{b\beta c\sigma} \bar{\rho}^{(2)b\beta c\sigma}_{mn} E^{b}_{\beta} E^{c}_{\sigma} e^{-i\omega_{\Sigma}t}
\label{eq:rhobar2nd}
\end{align}
where $\omega_{\Sigma} \equiv \omega_{\beta} +  \omega_{\sigma}$.

\subsection{First order density matrix}
To linear order we find two physically distinct terms
\begin{align}
\bar{\rho}^{(1)}_{mn} &=  \bar{\rho}^{(1e)}_{mn} + \bar{\rho}^{(1i)}_{mn},
\label{eq:rho1st}
\end{align}
where the superscripts $e$ and $i$ indicate interband and intraband processes, respectively 
\begin{align}
\bar{\rho}^{(1e) b\beta}_{mn} &=\frac{e}{\hbar} \frac{r_{mn}^{b}f_{nm}}{\omega_{mn} -\bar{\omega}_{\beta}}, 
\label{eq:rho1st21} \\
\bar{\rho}^{(1i) b\beta}_{mn}  &=  \delta_{nm} \frac{e}{\hbar} \frac{f_{n;b}}{i\bar{\omega}_{\beta}}.
\label{eq:rho1st22}
\end{align}
Here $f_n= [e^{(\mathcal{E}_n-\mu)/k_B T}+1]^{-1}$ is the Fermi distribution function at temperate $T$ and chemical potential $\mu$, $\mathcal{E}_n=\mathcal{E}_{n}(\v{k})$ is the energy dispersion of Bloch electrons in band $n$, and $\bar{\omega}_{\beta} \equiv \omega_{\beta} + i/\tau$ (see also Appendix~\ref{sec:defs}). 

Note that the Hall conductivity is obtained from Eq.(\ref{eq:rho1st21}) in the dc and $\tau \to \infty$ limits, see Appendix \ref{sec:interbad_sigma1e}, and the Drude conductivity from Eq.(\ref{eq:rho1st22}). Only when both are taken into account we recover the full quantum mechanical conductivity to linear order, see Appendix \ref{sec:Drude_sigma_linear}.

\subsection{Second order density matrix}
\label{sec:rho_2nd}
To second order we obtain again two physically distinct terms
\begin{align}
\bar{\rho}^{(2)}_{mn} &= \bar{\rho}^{(2e)}_{mn}  + \bar{\rho}^{(2i)}_{mn},
\label{eq:rho2nd}
\end{align}
where the superscripts $e$ and $i$ indicate interband and intraband processes, respectively 
\begin{align}
\bar{\rho}^{(2e)b\beta c\sigma}_{mn}&=\frac{ie}{\hbar (\omega_{mn}- \bar{\omega}_{\Sigma})} \big[ \bar{\rho}_{mn;c}^{(1e)b\beta} \nn\\
&~~~~~+ i \sum_{l} (\bar{\rho}_{ml}^{(1e)b\beta}r_{ln}^{c} -  r_{ml}^{c}\bar{\rho}_{ln}^{(1e)b\beta})\big],
\label{eq:rho2e} \\
\bar{\rho}^{(2i)b\beta c\sigma}_{mn} &= \frac{e^2}{\hbar^2} \frac{1}{i\bar{\omega}_{\sigma}} \frac{r_{mn}^{b}f_{nm;c}}{\omega_{mn} -\bar{\omega}_{\Sigma}} + \delta_{nm}  \frac{e^2}{\hbar^2} \frac{1}{i\bar{\omega}_{\sigma}}\frac{f_{n;cb}}{i\bar{\omega}_{\Sigma}}.
\label{eq:rho2i}
\end{align}
We defined $\bar{\omega}_{\beta} \equiv \omega_{\beta} + i/\tau,~ \bar{\omega}_{\Sigma} \equiv \omega_{\Sigma} + i/\tau$, and $\omega_{\Sigma} \equiv \omega_{\beta} +\omega_{\sigma}$. The terms in Eq.(\ref{eq:rho2i}) are FS contributions, since they contain derivatives of the distribution function $f_n$. In particular, the first term in Eq.(\ref{eq:rho2i}) involves transitions from the FS to other bands.\cite{Gao2021,Mahon2021}  FS contributions to higher orders can be computed in a similar way. For example, to $nth$ order there will always be a FS contribution proportional to the $nth$ derivative of $f_n$.

\section{Ground state spin polarization}
Each Bloch electron contributes
\begin{align}
\v{s}_{nm} =  \frac{\hbar}{2} \langle n\v{k} |\pmb{\sigma} |m\v{k}\rangle
\end{align}
to the total spin expectation value
\begin{align}
\v{S}&=\frac{1}{V}\sum_{nm\v{k}} \rho_{mn} \v{s}_{nm}.
\end{align}
Here $n=(n i_{n})$ labels band index $n$ and spinor index $i_{n}=1,2$, and $\pmb{\sigma}=(\sigma_x,\sigma_y,\sigma_z)$ are the Pauli spin matrices. In the ground state $\rho^{(0)}_{mn}=\delta_{mn}f_n$, and the spin magnetization is 
\begin{align}
\v{S}_0&=\frac{1}{V}\sum_{n\v{k}} f_{n} \v{s}_{n},
\end{align}
where we defined $\v{s}_{n} \equiv \v{s}_{nn}$. If there is TRS, we can choose $\v{s}_{nm}(-\v{k})= -\v{s}_{\bar{m}\bar{n}}(\v{k})$, $f(\mathcal{E}_{n}(-\v{k}))=f(\mathcal{E}_{\bar{n}}(\v{k})) $, and hence $\v{S}_0=0$, as expected. See also Appendix~\ref{sec:symmetry}. $\bar{n}$ is the spin-flipped state.

\section{First order spin polarization: spin Edelstein effect}
\label{sec:zeta1_gen}
To first order the induced spin 
\begin{align}
S^{(1)a} &=\sum_{b\beta} \zeta^{(1)ab}(-\omega_{\beta};\omega_{\beta}) E_{\beta}^{b} e^{-i\omega_{\beta}t},
\label{eq:S1_gen}
\end{align}
oscillates at the frequency of the external field. It is convenient to analyze the Fermi surface (FS) and non-FS (nFS) contributions separately. To this end, we write
\begin{align}
\zeta^{(1)} &= \zeta_{nFS}^{(1)}+ \zeta_{FS}^{(1)}
\label{eq:zeta1gen}
\end{align}
where 
\begin{align}
\zeta_{nFS}^{(1)ab} &= \frac{e}{\hbar V} \sum_{nm\v{k}} \frac{f_{nm} r_{mn}^b s_{nm}^{a}}{\omega_{mn}-\bar{\omega}_{\beta}} 
\label{eq:1inter}
\\
\zeta_{FS}^{(1)ab} &= \frac{e}{\hbar V} \frac{1}{i\bar{\omega}_{\beta}} \sum_{n\v{k}} f_{n;b} s_{n}^{a}.
\label{eq:1FS}  
\end{align}
If there is IS, the FS and nFS contributions vanish.  To see this, let $\v{k} \to -\v{k}$ in the integrands, and note that we can choose $\v{r}_{nm}(-\v{k})= -\v{r}_{nm}(\v{k})$ if there is IS. The spin texture at the FS determines $\zeta_{FS}^{(1)}$, but an integration by parts shows that we can also think of Eq.(\ref{eq:1FS}) as the average \textit{spin dipole} (momentum derivative of the spin) of the Fermi sea
\begin{align}
\zeta_{FS}^{(1)ab} &= -\frac{e}{\hbar V} \frac{1}{i\bar{\omega}_{\beta}} \sum_{n\v{k}} f_{n} s_{n;b}^{a}.
\label{eq:zeta1FS_alternative} 
\end{align}
In general, an inhomogeneous spin texture in momentum space generates a spin polarization in real space.

\section{Special cases of the linear spin susceptibility}
\label{sec:zeta1_special_cases}
Eqs.(\ref{eq:1inter}) and (\ref{eq:1FS}) can be specialized to any frequency configuration of external fields. For example, for a monocromatic field with components $\omega_{\beta}=\pm \omega$ or a static field we have
\begin{align}
\zeta^{(1)}_{ac} &\equiv \zeta^{(1)}(-\omega,\omega), \\
\zeta^{(1)}_{e} &\equiv \zeta^{(1)}(0,0),
\end{align}
which are the well-known ac and dc Edelstein spin susceptibilities. Explicitly
\begin{align}
\zeta_{ac}^{(1)ab} &=\frac{e}{\hbar V}\sum_{nm\v{k}} \frac{f_{nm} r_{mn}^{b} s_{nm}^{a}}{\omega_{mn}-\bar{\omega}},
\label{eq:zeta1_ac} \\
\zeta_{e}^{(1)ab} &= -\frac{e\tau }{\hbar V} \sum_{n\v{k} } f_{n;b} s_{n}^{a}.
\label{eq:zeta1_dc}
\end{align}
To obtain Eq.(\ref{eq:zeta1_dc}) we assumed TRS and large $\tau$. These expressions agree with, e.g., Kubo formula results.\cite{Freimuth2014, Zelezny2017,Gustav2017}

\begin{figure}[]
\subfigure{\includegraphics[width=.48\textwidth]{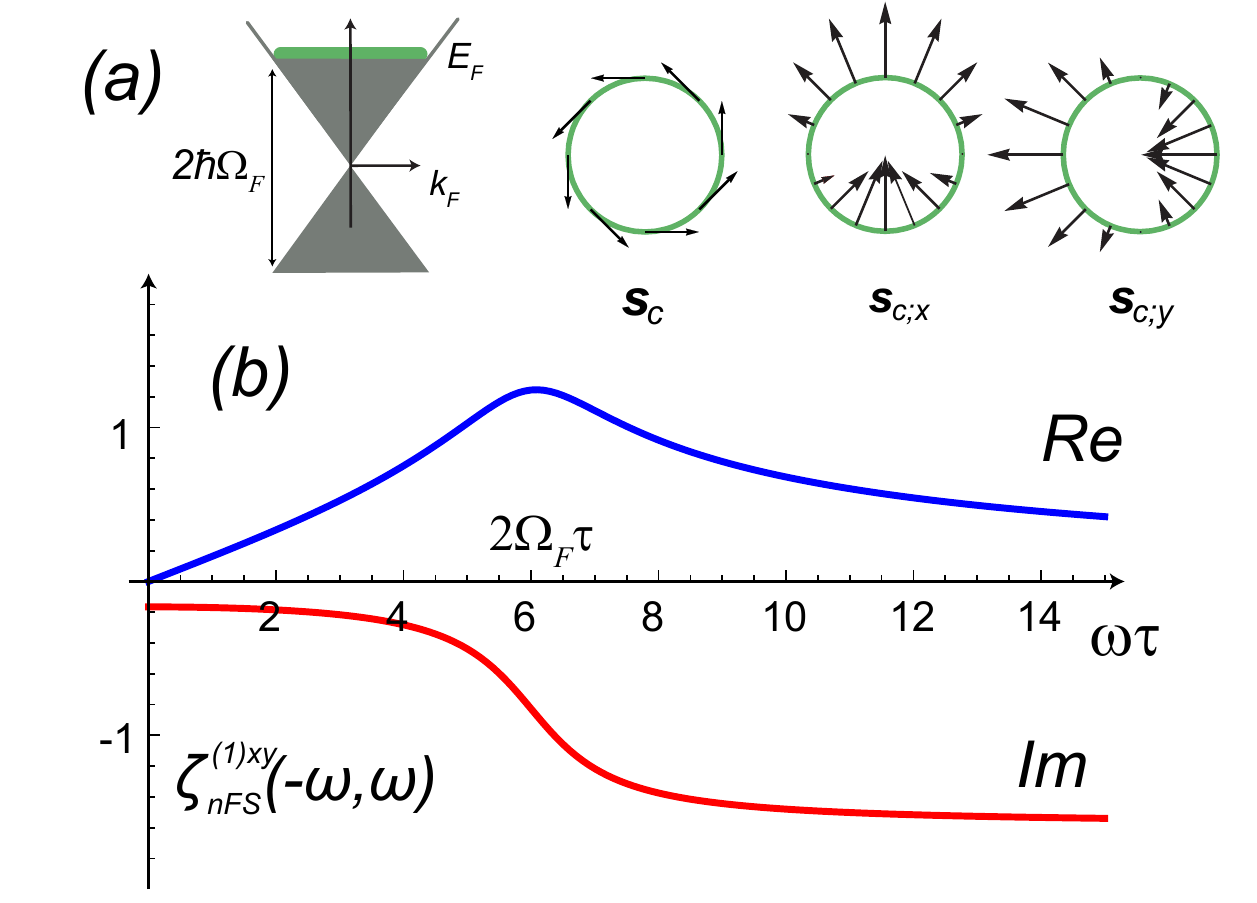}}
\caption{(a) Spin textures of conduction band $\v{s}_c \sim \hat{\v{z}}\times \hat{\v{k}}$ and spin dipoles at the Fermi surface. (b) Non-FS contribution to electric spin susceptibility in TIs in units of ($e/16\pi v$)  and for $\Omega_F\tau=3$.}
\label{fig:spin_texture}
\end{figure}

\section{Example: TI in an electric field}
\label{sec:zeta1ej}
Consider the electrons at the surface of the topological insulator (TI).\cite{Moessner2021} Such electrons have, in a sense, maximal SOC. The spin forms a vortex (antivortex)  in the conduction (valence) band with a center at Dirac point $\v{k}=0$. The conduction and valence bands have conical shape with apices meeting at the Dirac point, Fig.~\ref{fig:spin_texture}(a). An effective Hamiltonian near the Dirac point is
\begin{align}
H_{TI}= v \hbar (k^{x} \sigma_{y} - k^{y} \sigma_{x}),
\end{align}
where $v$ is the slope of quasiparticles. Let us assume the Fermi level $\mathcal{E}_{F}$ lies in the conduction band, Fig.~\ref{fig:spin_texture}(a). For a monocromatic linearly polarized light incident perpendicular to the surface $\v{E} = 2^{-1}\v{E}_0 e^{-i\omega t} + c.c.$ of frequency $\omega$, Eq.(\ref{eq:S1_gen}) becomes 
\begin{align}
\v{S}^{(1)} = \v{S}^{(1)}_{nFS} + \v{S}^{(1)}_{FS},
\label{eq:spin_TI}
\end{align}
where
\begin{align}
\v{S}^{(1)}_{nFS} &= S_{ti} \textrm{ArcCot}\left(\frac{2\Omega_F \tau}{1-i \omega\tau} \right) (\hat{\v{z}}\times \hat{\v{E}}_0) e^{i\omega t} + c.c,
\label{eq:S1nFS}\\
\v{S}^{(1)}_{FS} &= \frac{S_{ti} (\hat{\v{z}}\times \hat{\v{E}}_0)  ~4 \Omega_F \tau}{\omega^2\tau^2 +1} (\omega\tau\sin(\omega t) + \cos(\omega t)).
\label{eq:S1FS}
\end{align}
 We defined $\v{E}_0= E_{0} \hat{\v{E}}_0$ and $\Omega_F \equiv v k_F$. The order of magnitude of the spin polarization is given by $S_{ti}\equiv eE_0/32\pi v$ and has units of spin/m$^2$. Note that only the $xy$ components of the susceptibilities are nonzero.  The nFS spin susceptibility is shown in Fig.~\ref{fig:spin_texture}(b). Its real part has a maximum  at $\omega= 2\Omega_F$, i.e., as interband transitions become possible. Its imaginary part has a step-like feature at $\omega= 2\Omega_F$ as energy absorption becomes favorable. The FS contribution to the spin susceptibility has a decaying behavior as a function of frequency, similar to the Drude conductivity. Note that both the FS and nFS spin components depend on the dimensionless parameters $\Omega_F \tau$, which parametrize the density/cleanness of the surface.

Interestingly, the spin polarization is perpendicular to the electric field. Intuitively,\cite{Culcer2010,Pesin2012}  the spin and velocity are related by $\v{s}_{c}= (\hbar/2v) \hat{\v{z}}\times \v{v}_{c}$, where $c$ labels the conduction band, and hence when the electronic velocity obtains a nonzero expectation value (in the direction of the electric field), so does the spin polarization (perpendicular to it). Eq.(\ref{eq:zeta1FS_alternative}) makes this statement more precise because the spin polarization is also the average spin dipole 
\begin{align}
\v{s}_{c;x} &= \frac{\hbar  \hat{\v{k}}}{2 k }\sin\theta, 
\label{eq:scx_scy1} \\ 
\v{s}_{c;y} &= -\frac{\hbar \hat{\v{k}}}{2 k }\cos\theta, 
\label{eq:scx_scy2}
\end{align} 
over the Fermi sea, Fig.~\ref{fig:spin_texture}(a). Here, $\hat{\v{k}}=(\cos\theta,\sin\theta)$ is the unit vector along momentum and $\theta$ is the angle of $\v{k}$ with the $x$-axis. The spin vortex texture in this specific case gives a transverse spin polarization. Finally, with a static electric field, only the FS term contributes ($\Omega_F \tau \gg 1$) and we obtain
\begin{align}
\v{S}^{(1)}_{dc} = S_{ti}~ 4\Omega_F \tau  (\hat{\v{z}}\times \hat{\v{E}}_0).
\end{align}
A significant spin polarization ratio could be achieved. For example, if $E_0= 10^{4}$ V/m and $v= 10^{6}$ m/s we obtain $ 2S_{ti}/\hbar \sim 5\times 10^4$ Borh magnetons per mm$^2$, i.e., roughly the equivalent of $5\times 10^4$ fully polarized electrons per mm$^{2}$. Typical surface electron density is $10^{10}$ mm$^{-2}$. If the density can be brought to, e.g., $10^{6}$ mm$^{-2}$ or less, the spin polarization would be $5$\% or more.

\section{Second order spin polarization} 
\label{sec:zeta2_general} 
To second order, the induced spin 
\begin{align}
S^{(2)a} &= \sum_{b\beta c \sigma} \zeta^{(2)abc}(-\omega_{\Sigma},\omega_{\beta},\omega_{\sigma}) E^{b}_{\beta} E^{c}_{\sigma} e^{-i\omega_{\Sigma} t},
\label{eq:S2_gen}
\end{align}
oscillates at the frequency $\omega_{\Sigma} = \omega_{\beta} + \omega_{\sigma}$ in the long-time limit. It has two contributions, which we label as intraband and interband 
\begin{align}
\zeta^{(2)}= \zeta^{(2e)} + \zeta^{(2i)},
\label{eq:zeta2nd_inter_intra}
\end{align}
because they arise from Eq.(\ref{eq:rho1st21}) and Eq.(\ref{eq:rho1st22}), respectively.

\subsection{Second order intraband contribution} 
To second order, Eqs.(\ref{eq:rho1st21}) and (\ref{eq:rho1st22}) branch out to produce interband and intraband terms when substituted into the Boltzmann equation. Let us dub second-order intraband those terms which originate from Eq.(\ref{eq:rho1st22}). From Eq.(\ref{eq:rho2i}) we can separate the 2nd order intraband terms further into interband and intraband processes as 
\begin{align}
\zeta^{(2i)}= \zeta^{(2ie)} + \zeta^{(2ii)}, 
\end{align}
where 
\begin{align}
\zeta^{(2ie)abc}&= \frac{e^2}{\hbar^2} \frac{1}{i\bar{\omega}_{\sigma}} \frac{1}{V}\sum_{nm\v{k}}   \frac{s_{nm}^{a} r^{b}_{mn} f_{nm;c}}{ \omega_{mn}-\bar{\omega}_{\Sigma}},
\label{eq:chi2Fse}\\
\zeta^{(2ii)abc}&= \frac{e^2}{\hbar^2} \frac{1}{i\bar{\omega}_{\sigma} i\bar{\omega}_{\Sigma}} \frac{1}{V}\sum_{n\v{k}} s_{n}^{a} f_{n;cb}. 
\label{eq:2FSi}
\end{align}
These expressions still need to be symmetrized with respect to the exchange of indices $\beta b \leftrightarrow \sigma c$. Note that $\zeta^{(2ie)abc}$ is proportional to $f_{nm;c}\equiv f_{n;c}-f_{m;c}$ which probes transitions from electrons at a FS(s) to higher/lower energy bands. Interestingly, $\zeta^{(2ii)}$ is proportional to the spin quadrupole moment which vanishes if there is TRS, but does not necessarily vanish if there is IS.

\subsection{Second order interband contribution}
\label{sec:zeta2ndnFS} 
The interband contribution to second order is 
\begin{align}
\zeta^{(2e)abc}&=\frac{ie}{\hbar V} \sum_{nm\v{k}} \frac{s_{nm}^{a}}{\omega_{mn}- \bar{\omega}_{\Sigma}}
\bigg[\bar{\rho}_{mn;c}^{(1e)b\beta} \nn\\
&~~~~~~~~~+ i \sum_{l} (\bar{\rho}_{ml}^{(1e)b\beta}r_{ln}^{c} -  r_{ml}^{c}\bar{\rho}_{ln}^{(1e)b\beta})\bigg].
\label{eq:zeta2e} 
\end{align}
This expression still needs to be symmetrized with respect to exchange $b \beta \leftrightarrow c\sigma$.

\section{Special cases of the second order spin susceptibility} 
\label{sec:zeta2_special_cases}
Eq.(\ref{eq:S2_gen}) can be specialized to particular configurations of fields. For example, we can construct quadratic susceptibilities 
\begin{align}
\zeta^{(2)}_{e} &\equiv \zeta^{(2)}(0,0,0), 
\label{eq:zeta2000} \\
\zeta^{(2)}_{bpse} &\equiv \zeta^{(2)}(0,\omega,-\omega), 
\label{eq:zetabpse} \\
\zeta^{(2)}_{e,ac} &\equiv \zeta^{(2)}(-\omega,0,-\omega),
\label{eq:zetamix} \\
\zeta^{(2)}_{shg} &\equiv \zeta^{(2)}(-2\omega,\omega,\omega),
\label{eq:zetashg}
\end{align}
with an optical monocromatic source and a dc static field. Eq.(\ref{eq:zeta2000}) is the dc quadratic correction to the Edelstein susceptibility shown in Eq.(\ref{eq:S_Elec_exp}) (see Sec.~\ref{sec:zeta2e000}).  Eq.(\ref{eq:zetabpse}) is the bulk photospin susceptibility (see Sec.~\ref{sec:bpse_ins} and \ref{sec:bpse_met}). Eq.(\ref{eq:zetamix}) corresponds to the Edelstein susceptibility modulated by an ac field. Eq.(\ref{eq:zetashg}) is the generation of second harmonics in the spin polarization.

\section{Quadratic Edelstein susceptibility}
\label{sec:zeta2e000}
Direct calculation of Eq.(\ref{eq:zeta2000}) yields intraband and interband terms. However, if we assume TRS, all terms vanish except for the FS contribution Eq.(\ref{eq:chi2Fse}) and we obtain
\begin{align}
\zeta^{(2)abc}_{e}& = -\frac{e^2\tau}{2\hbar^2 V} \sum_{nm\v{k}} \frac{s_{nm}^a}{\omega_{mn}}\{ r_{mn}^b, f_{nm;c}\}. 
\label{eq:zeta2dc}
\end{align}
See Appendix~\ref{sec:defs} for notation and Appendix~\ref{sec:symmetry} for more details on symmetry constrains. Note  that Eq.(\ref{eq:zeta2dc}) is real and symmetric under exchange of the indices $b \leftrightarrow c$. From this result we conclude that there is no second-order dc Edelstein susceptibility in insulators.

\section{Bulk photospin effect in insulators}
\label{sec:bpse_ins}
We now consider an insulator with fully occupied valence bands and fully empty conduction bands so that Eq.(\ref{eq:chi2Fse}) and Eq.(\ref{eq:2FSi}) vanish. Let us assume there is a monochromatic optical field of the form $\v{E} = \v{E}(\omega) e^{-i\omega t}+c.c.$. The \textit{static} induced spin is
\begin{align}
S^{(2)a}_{bpse,in}= 2 \sum_{bc} \zeta^{(2e)abc}(0;\omega,-\omega) E^{b}(\omega) E^{c}(-\omega).
\label{S2ndnFSexample}
\end{align}
We now separate the symmetric and antisymmetric parts of the \textit{interband} response by defining  
\begin{align}
\nu_{2,in}^{abc} &\equiv (\zeta^{(2e)abc} + \zeta^{(2e)acb})/2, 
\label{eq:nu_upsilon_defs1} \\
\upsilon_{2,in}^{abc} &\equiv (\zeta^{(2e)abc} - \zeta^{(2e)acb})/2.
\label{eq:nu_upsilon_defs2}
\end{align}
In term of these Eqn.~\ref{S2ndnFSexample} becomes
\begin{align}
S^{(2)a}_{bpse,in}= 2 \sum_{bc} \nu_{2,in}^{abc}& E^{b}(\omega) E^{c}(-\omega)\nn \\
&+ 2 \sum_{bc} \upsilon_{2,in}^{abc} E^{b}(\omega) E^{c}(-\omega),
\label{S2ndnFSexample2}
\end{align}
which is of the form of Eqn. \ref{eq:Sdc_s_as}. From Eq.(\ref{eq:zeta2e}), $\zeta^{(2e)abc}$ can be further decomposed into 2-band and 3-band contributions 
\begin{align}
\zeta^{(2e)}= \zeta_{2b}^{(2e)} +\zeta_{3b}^{(2e)},
\end{align}
where
\begin{align}
\zeta_{2b}^{(2e)abc} &= -\frac{ie^2}{2\hbar^2 V}\sum_{nm\v{k}} \frac{s_{nm}^a }{\bar{\omega}_{nm}} \bigg[ \left(\frac{r_{mn}^b f_{nm}}{\omega_{mn}-\bar{\omega}}\right)_{;c}  \nn \\
&~~~~~~~~~~~~~~~~~~~~~~~~~ + \left(\frac{r_{mn}^c f_{nm}}{\omega_{mn}+\bar{\omega}^{*}}\right)_{;b}  \bigg],
\label{eq:zeta2nFS2b} \\
\zeta_{3b}^{(2e)abc} &= \frac{e^2}{2\hbar^2 V}\sum_{nml\v{k}} \frac{s_{nm}^a}{\bar{\omega}_{nm}} \bigg[ \frac{r_{ml}^b r_{ln}^c f_{lm}}{\omega_{ml}-\bar{\omega}} - \frac{r_{ml}^c r_{ln}^b f_{nl}}{\omega_{ln}-\bar{\omega}} \nn \\ 
&~~~~~~~~~~~~~~~~~+ \frac{r_{ml}^c r_{ln}^b f_{lm}}{\omega_{ml}+\bar{\omega}^{*}} - \frac{r_{ml}^b r_{ln}^c f_{nl}}{\omega_{ln}+\bar{\omega}^{*}}\bigg],
\label{eq:zeta2nFS3b}
\end{align}
and $\bar{\omega} \equiv \omega + i/\tau$ and $\bar{\omega}_{nm} = \omega_{nm} + i/\tau$. 

Eqs.(\ref{eq:zeta2nFS2b}) and (\ref{eq:zeta2nFS3b}) are the most general for any $\tau$. However, in the limit of large $\tau$, we can set $\bar{\omega}_{nm} = \omega_{nm} + i/\tau  \to \omega_{nm}$ when $n\neq m$ in denominators. In this case, many terms simplify if there is TRS. For example, the symmetric part becomes
\begin{align}
\nu_{2,in}^{abc} &= -\frac{\pi e^2}{2\hbar^2 V}\sum_{nm\v{k}} f_{nm}\{\left( \hspace{-2pt} \frac{s^{a}_{nm}}{\omega_{nm}}\hspace{-2pt} \right)_{;c} \hspace{-2pt},  r_{mn}^{b} \} \delta_{\tau}(\omega_{mn}-\omega) \nn \\
&~~~+ \frac{i \pi e^2}{2\hbar^2 V}\sum_{
\begin{smallmatrix}
nml\v{k} \\
n\neq m 
\end{smallmatrix}
} \left( \hspace{-2pt} \frac{s^{a}_{nm}}{\omega_{nm}}\hspace{-2pt} \right) f_{lm} \{ r_{ml}^{b}, r_{ln}^{c} \} D_{+}(\omega_{ml},\omega),
\label{eq:sigma2}
\end{align}
which is real and resonant, i.e., vanishes for field frequencies smaller than the energy gap. The notation is defined in Appendix~\ref{sec:defs}. Note that $\nu_{2,in}$ depends on the off-diagonal elements of the density matrix and hence is a pure quantum effect. In particular, the first term in Eq.(\ref{eq:sigma2}) comes from Eq.(\ref{eq:zeta2nFS2b}) whose $m=n$ term vanishes. The second term in Eq.(\ref{eq:sigma2}) comes from Eq.(\ref{eq:zeta2nFS3b}) with the $n=m$ term excluded. The $n=m$ term of Eq.(\ref{eq:zeta2nFS3b}) in fact is the resonant part of $\upsilon_{2,in}$
\begin{align}
\upsilon_{2,in}^{abc} &=  \frac{\pi e^2 \tau}{2\hbar^2 V}\sum_{nm\v{k}} (s^{a}_n - s^{a}_{m}) f_{mn} [r_{nm}^{b}, r_{mn}^{c} ] \delta_{\tau}(\omega_{nm}-\omega) \nn \\
&~~+\frac{i e^2}{4\hbar^2 V}\sum_{nm\v{k}} f_{nm} [ \left( \hspace{-2pt} \frac{s^{a}_{nm}}{\omega_{nm}}\hspace{-2pt} \right)_{;c} \hspace{-2pt},  r_{mn}^{b} ] H_{-}(\omega_{mn},\omega) \nn \\
& ~~+  \frac{ e^2}{2\hbar^2 V}\sum_{
\begin{smallmatrix}
nml\v{k} \\
n\neq m 
\end{smallmatrix}
} \left(\hspace{-2pt} \frac{s^{a}_{nm}}{\omega_{nm}}\hspace{-2pt} \right) f_{lm} [ r_{ml}^{b}, r_{ln}^{c} ] H_{-}(\omega_{ml},\omega).
\label{eq:eta2}
\end{align}
$\upsilon_{2,in}$ is pure imaginary and contains both resonant and nonresonant contributions. The resonant part is proportional to $\tau$, meaning a spin injection can be obtained from simple Fermi's golden applied to the diagonal elements of the density matrix, e.g., Eq.(4) of Ref.~\onlinecite{Nastos2007}. Alternatively, the resonant term of $\upsilon_{2,in}$ could have been derived from the effective equation of motion 
\begin{align}
\frac{d}{d t}\v{S}^{(2)} = \left(\frac{d }{dt} \v{S}^{(2)}\right)_{source} - \frac{1}{\tau}\v{S}^{(2)},
\end{align}
with recombination and spin relaxation times equal to $\tau$.~\cite{Meier1984,Zutic2004}  The resonant part of $\upsilon_{2,in}$ has an intuitive physical explanation: As the electron absorbs the energy of a photon and jumps from a valence to a conduction band, the angular momentum of the photon transfers to the electron spin $\Delta \v{s}= \v{s}_c - \v{s}_v$.

Interestingly, the last two terms in Eq.(\ref{eq:eta2}) are nonresonant; i.e., they are nonzero even for subgap frequencies. This means a permanent spin polarization is possible with circularly polarized photons even if energy is not absorbed and heat is not produced. This is important for spintronic applications. However, a more detailed model of dissipation is needed to understand the evolution of interband coherence. 

\section{Bulk photospin effect in metals} 
\label{sec:bpse_met}
The static spin response of metals includes both intraband and interband contributions, see Eqn.~\ref{eq:zeta2nd_inter_intra} 
\begin{align}
S^{(2)a}_{bpse,m}= 2 \sum_{bc} \zeta^{(2)abc}(0;\omega,-\omega) E^{b}(\omega) E^{c}(-\omega),
\label{S2nd_metals}
\end{align}
Accordingly, we now define the BPSE response tensor for metals as the symmetric and antisymmetric $\zeta^{(2)abc}$
\begin{align}
\nu_{2,m}^{abc} &\equiv (\zeta^{(2)abc} + \zeta^{(2)acb})/2,  \\
\upsilon_{2,m}^{abc} &\equiv (\zeta^{(2)abc} - \zeta^{(2)acb})/2.
\end{align}
In addition to the interband contributions of insulators, metals have FS-specific contributions.  Assuming TRS, we find that only Eq.(\ref{eq:chi2Fse}) survives, and

\begin{align}
\nu_{2,m} &= \nu_{2,in} + \nu_{2,FS}, \\ 
\upsilon_{2,m} &= \upsilon_{2,in} + \upsilon_{2,FS},
\label{eq:sigma2and eta2metal}
\end{align}
where the metallic contributions are 
\begin{align}
\nu_{2,FS}^{abc} &= -\frac{\tau e^2}{2\hbar^2 V}\frac{1}{1+ \omega^{2}\tau^2} \sum_{nm\v{k}} 
\left( \hspace{-2pt} \frac{s^{a}_{nm}}{\omega_{mn}}\hspace{-2pt} \right) \hspace{-2pt} \{ r_{mn}^{b},f_{nm;c} \} \\
\upsilon_{2,FS}^{abc} &= \frac{i \tau e^2}{2\hbar^2 V}\frac{\omega\tau}{1+ \omega^{2}\tau^2} \sum_{nm\v{k}} 
\left( \hspace{-2pt} \frac{s^{a}_{nm}}{\omega_{mn}}\hspace{-2pt} \right) \hspace{-2pt} [ r_{mn}^{b},f_{nm;c} ].
\label{eq:sigma2metal22}
\end{align}

\section{Example: TI in an electric and magnetic field}
\label{sec:example_ti_mag_elec}
We consider the electrons at the surface of a TI subject to an Zeeman field 
\begin{align}
H= v \hbar (k^x \sigma_{y} - k^y \sigma_{x}) + m\sigma_{z}.
\end{align}
The Zeeman field breaks TRS and now the spin quadrupole Eq.(\ref{eq:2FSi}) does not vanish. Let us assume an electric field linearly polarized in the plane of the TI surface with magnitude $E_0$ and frequency $\omega$. The induced spin points out of the plane and has magnitude  
\begin{align}
S^{(2ii)z} = -\frac{e^2 \tau^2 E_0^2}{4\pi \hbar^2}\frac{s_{c}^{z}(k_F)}{\omega^2\tau^2 + 1} \left(1- \frac{m^2}{\mathcal{E}_F^2}\right),
\end{align} 
where $s_{c}^{z} = \hbar m/ 2\mathcal{E}_c$ is the z-component of the spin in the conduction band and $\mathcal{E}_F\equiv \mathcal{E}_c(k_F) >m$ is the Fermi level assumed to lie in the conduction band.  If $\mathcal{E}_F\gg m$ and the electric field is static, the spin quadrupole is given by 
\begin{align}
S^{(2)z}_{quad} = -\frac{e^2 \tau^2 E_0^2 }{4\pi \hbar^2} s_{c}^{z}(k_F)  ~~~~~~\omega=0.
\end{align} 

\section{Example: Coherence of spin-split conduction bands} 
\label{sec:opical_inj_ex}
At very short times, interband coherence of conduction bands plays an important role in insulators that break IS.\cite{Bhat2005,Nastos2007,Nastos2010,Rioux2012,ZapataPena2017} Here we recover the equations describing this effect starting from Eq.(\ref{eq:zeta2e}).  To include the coherence of pairs of conduction bands close in energy, instead of taking the diagonal elements of Eq.(\ref{eq:zeta2nFS2b}) (which give zero) and Eq.(\ref{eq:zeta2nFS3b}), we consider the diagonal elements of Eq.~\ref{eq:zeta2nFS2b} and the first and fourth terms of Eq.(\ref{eq:zeta2nFS3b})
\begin{align}
\bar{\zeta}_{ch}^{(2)abc} = \frac{e^2}{2\hbar^2 V} \sum_{nml\v{k}} \frac{s_{nm}^a r_{ml}^b r_{ln}^c}{\bar{\omega}_{nm}} [\frac{f_{lm}}{\omega_{ml}-\bar{\omega}} -  \frac{f_{ln}}{\omega_{nl}-\bar{\omega}^{*}}],
\end{align}
and let $n=c'$ and $m=c$ label conduction bands spin-split by SOC but very close in energy.  Now write $\omega_{c' v}= \omega_{c v} - \omega_{cc'}$ and expand in powers of the small parameter $\omega_{cc'}$. To lowest order, we obtain
\begin{align}
\bar{\zeta}_{ch}^{(2)abc}\bigg|_{\begin{smallmatrix}
n=c',  \\
m=c\phantom{-}
\end{smallmatrix}}&= \frac{\tau \pi e^2}{2\hbar^2 V} \sum_{cc'\v{k}} s_{cc'}^a r_{c'v}^b r_{vc}^c[ \delta_{\tau}(\omega_{c'v}-\omega) \nn \\
&~~~~~~~~~~~~~~~~~~~~~~~+ \delta_{\tau}(\omega_{cv}-\omega) ],
\label{eq:zeta2nFS_nastos}
\end{align}
which was originally derived in Ref.~\onlinecite{Nastos2007} by other methods.

\section{Discussion and conclusions}
\label{sec:conclusion}
We calculated the electric spin susceptibility to second order in the electric field for a system of Bloch electrons with SOC. We dub this response a \textit{bulk photospin effect} (BPSE) to emphasize that a spin polarization is generated in the bulk of the materials without need of interfaces. Our expressions for the BPSE tensors depend on generic Bloch matrix elements and hence are amenable for use in large-scale first-principles numerical codes. In appropriate limits, we recover the linear Edelstein and spin orientation phenomena. 

We can draw some general conclusions from the form of the BPSE susceptibility: (a) The symmetric part of the BPSE susceptibility in insulators is resonant; i.e., light of linear polarization cannot induce spin polarization unless the frequency of light is at least equal to the energy gap, and hence energy is absorbed by the electron ensemble. 

(b) The symmetric part of the BPSE susceptibility in insulators depends on the off-diagonal elements of the density matrix; i.e., spin polarization with linearly polarized light requires quantum coherence. Since quantum coherence can be feeble, this kind of spin polarization may be harder to observe in experiments.

(c) The antisymmetric part of the BPSE susceptibility in insulators vanishes for linearly polarized light and is maximum for circular polarization. In this sense the antisymmetric part of the BPSE susceptibility characterizes the response of the system to the chirality of light and represents an instance of the inverse Faraday effect. The antisymmetric part of the BPSE susceptibility has both resonant and nonresonant contributions. The former is given by diagonal elements of the density matrix and reproduces the standard spin orientation effects which requires energy absorption. The latter, means a spin polarization is possible with circularly polarized light even if the system does not absorbs energy. 

(d) Non-magnetic metals have additional contributions to the spin polarization. The symmetric and antisymmetric parts of the metallic response are nonzero at all frequencies and hence produce heating. Although linear or circular polarization of light can induce spin polarization this requires the existence quantum coherence. 
 
We have seen that, to linear order, the FS contribution to the static spin polarization is proportional to the average spin dipole moment of the Fermi sea. To second order the FS contribution to the static spin polarization is proportional to the average spin quadrupole of the Fermi sea. In general, the FS contribution to the $m$-th order electric spin susceptibility can be thought, as the average $m$-th spin multipole of the Fermi sea
\begin{align}
\zeta_{e}^{(m)abc...} = \left(\frac{e\tau}{\hbar}\right)^{m} \frac{1}{V} \sum_{n\v{k}} f_{n} s^{a}_{n;bc	...}.
\label{eq:spinnFS_dc_summ}
\end{align}
These terms are analogous to higher order Drude conductivities (see Appendix~\ref{sec:drude_2nd}) which are, so to speak, velocity moments of the Fermi sea
\begin{align}
\sigma_{dc}^{(1)ab} &= -\frac{e^2\tau}{\hbar V}  \sum_{n\v{k}} f_{n} v_{n;b}^{a}, \\
\sigma_{dc}^{(2)abc} &= -\frac{e^3\tau^2}{\hbar^2 V}  \sum_{n\v{k}} f_{n} v_{n;bc}^{a},
\label{eq:sigma1_2FS}
\end{align}

\section*{Acknowledgments}
We thank Bernardo S. Mendoza for useful discussions during the early stages of the project. We acknowledge support from NSF grant DMR-2015639 and DOE under contract DE-AC02-05CH11231 using NERSC award BES-ERCAP20386.

\appendix

\section{Definitions}
\label{sec:defs}
\begin{align}
e &=-|e| \\
\hat{H} &= \frac{\hat{p}^2}{2m} + V(\v{r}) + \mu_{B}^2 \v{e}\cdot (\hat{\v{p}}\times \pmb{\sigma}) 
\label{eq:hamiltonian} \\
\pmb{\sigma} &= (\sigma_x,\sigma_y,\sigma_z) ~~~ \textrm{Pauli spin matrices} \\
\langle \v{r}|n\v{k}\rangle &= \langle \v{r}|u_n\rangle e^{i\v{k}\cdot\v{r}} ~ \textrm{Bloch state ($n\v{k}$)},  \\
\mathcal{E}_n &= \langle n\v{k}| \hat{H}|n\v{k}\rangle,~~ \textrm{energy of state ($n\v{k}$)} \\
\hbar \omega_{n} &\equiv \mathcal{E}_n     \\
\rho_{mn} &= \langle a^{\dagger}_{n} a_{m} \rangle,~~ \textrm{density matrix Bloch basis} \\
f_n &= f(\mathcal{E}_n) = \rho_{nn}^{(0)}, ~~~~\textrm{Fermi function}  \\
\omega_{nm} &\equiv \omega_{n} - \omega_{m}  \\
f_{nm} &\equiv f_n - f_m  \\
\pmb{\xi}_{nm} &= \langle u_{n} | i\pmb{\nabla}_{\v{k}} | u_m \rangle,~~\textrm{Berry connection}  \\
\v{r}_{nm} &\equiv (1-\delta_{nm}) \pmb{\xi}_{nm} \\ 
\v{v}_{nm} &= \langle n\v{k} | \hat{\v{v}}| m\v{k} \rangle  \\
\rho_{mn;b} &\equiv \big[ \frac{\partial}{\partial k_b} - i (\xi_{mm}^{b} - \xi_{nn}^{b} )\big] \rho_{mn} \\
\v{r}_{mn;b} &\equiv \big[ \frac{\partial}{\partial k_b} - i (\xi_{mm}^{b} - \xi_{nn}^{b} )\big] \v{r}_{mn} \\
\v{v}_{mn;b} &\equiv \big[ \frac{\partial}{\partial k_b} - i (\xi_{mm}^{b} - \xi_{nn}^{b} )\big] \v{v}_{mn} \\
f_{n;b} &= \frac{\partial f_n}{\partial k^b} \\
\omega_{n;b} & \equiv \frac{\partial \omega_n}{\partial k^b} = v^{b}_{nn}= v^{b}_{n} \\
f_{nm;b} &\equiv f_{n;b}  -  f_{m;b} \\
\omega_{nm;b} &\equiv \omega_{n;b}  -  \omega_{m;b} \\
\v{s}_{nm} &\equiv \frac{\hbar}{2} \langle n\v{k} |  \pmb{\sigma}| m\v{k} \rangle, \\
\v{s}_{n} &\equiv \v{s}_{nn},	~~~~ \textrm{spin of state ($n\v{k}$) } \\
\v{s}_{mn;b} &\equiv \big[ \frac{\partial}{\partial k_b} - i (\xi_{mm}^{b} - \xi_{nn}^{b} )\big] \v{s}_{mn} \\
\v{s}_{n;b} &= \frac{\partial \v{s}_{n}}{\partial k^{b}},\\
\tau &= ~~\textrm{appropriate relaxation time} \\
\omega_{\beta} &= \textrm{frequency component of E-field} \\
\omega_{\Sigma} &= \omega_{\beta} + \omega_{\sigma} +\cdots  \\
\bar{\omega}_{\beta} &= \omega_{\beta} + i/\tau \\
\Omega_{n}^{ab} &= \frac{\partial \xi_{nn}^{b}}{\partial k^{a}} - \frac{\partial \xi_{nn}^{a}}{\partial k^{b}}  \\
\pmb{\Omega}_{n} &= \pmb{\nabla}\times \pmb{\xi}_{nn} ~~~~ \textrm{Berry curvature}
\end{align}
For any scalar functions $f(b)$ and $g(c)$ of Cartesian indices $b$ and $c$, we defined the symmetric and antisymmetric combinations

\begin{align}
\{f(b),h(c)\} &\equiv f(b)h(c) + f(c)h(b),  \\
[f(b),h(c)] &\equiv f(b)h(c) - f(c)h(b).
\end{align}
We also defined broadened delta functions as 

\begin{align}
\delta_{\tau}(x) &= \frac{1}{\pi} \frac{\tau^{-1}}{x^2 + \tau^{-2}},  \\
\frac{P_{\tau}}{x} &= \frac{x}{x^2 + \tau^{-2}},  \\
H_{\pm} (\omega_{mn},\omega) &\equiv \frac{P_{\tau}}{\omega_{mn}-\omega} \pm \frac{P_{\tau}}{\omega_{mn}+\omega},   \\
D_{\pm} (\omega_{mn},\omega) &\equiv \delta_{\tau}(\omega_{mn}-\omega) \pm \delta_{\tau}(\omega_{mn}+\omega).
\end{align}
$e$ is the charge of the electron, $\hat{\v{p}}=-i\hbar \pmb{\nabla}$ is the momentum operator, $V(\v{r})$ is the periodic ionic potential and $\v{e}$ the ionic spin-orbit field. 

\section{Useful relations}
An operator $\hat{Q}$ in Bloch basis can be projected into diagonal and off-diagonal components as
\begin{align}
\langle n\v{k}|\hat{Q}|m\v{k}\rangle = \delta_{nm} Q_{nn} + (1-\delta_{nm})Q_{nm}.
\end{align}
When $\hat{Q}$ is a commutator that involves the position matrix elements~\cite{Karplus1954,Blount1962}  
\begin{align}
\langle n\v{k}| \v{r} | m\v{k}'\rangle &= \delta_{nm}[\delta_{\v{k},\v{k}'}\pmb{\xi}_{nn} + i\pmb{\nabla}_{\v{k}}\delta_{\v{k},\v{k}'}] \nn \\
&~~~~~~~~~~~~~~~~~+ (1-\delta_{nm})\delta_{\v{k},\v{k}'}\pmb{\xi}_{nm},
\end{align}
the result are expressions that relate position, velocity, energy, current, and their (covariant) derivatives. The most common are 
\begin{align}
v^{b}_{nm} &= \delta_{nm} \omega_{n;b} +i \omega_{nm} r_{nm}^b  
\label{eq:v_sumr}\\
v^{b}_{nm;a} &= \frac{\hbar}{m}\delta_{ab}\delta_{nm} + i \sum_{l}(r_{nl}^{a} v_{lm}^{b} - v_{nl}^{b} r_{lm}^{a}) 
\label{eq:vnm_sumr} \\
\Omega_{n}^{ba} &= -i \sum_{l} ( r_{nl}^{a} r_{ln}^{b} - r_{nl}^{b} r_{ln}^{a}), 
\label{eq:v_Omeganba} \\
r^{a}_{nm;b} - r^{b}_{nm;a} &\dot{=} -i  \sum_{l}(r_{nl}^{a} r_{lm}^{b} - r_{nl}^{b} r_{lm}^{a}) 
\label{eq:rnma_rnmb_sumr} \\
\omega_{n;ab} &= \frac{\hbar}{m}\delta_{ab} - \sum_{l} \omega_{ln} (r_{nl}^{a} r_{ln}^b + r_{nl}^{b} r_{ln}^a) 
\label{eq:omega_nab}\\
r^{b}_{nm;a} &\dot{=} \frac{1}{\omega_{nm}} (r_{nm}^a \omega_{mn;b} + r_{nm}^{b}\omega_{mn;a}) \nn \\
& +  \frac{i}{\omega_{nm}} \sum_{l}(\omega_{lm} r_{nl}^{a} r_{lm}^{b} - \omega_{nl} r_{nl}^{b} r_{lm}^{a})
\label{eq:rnma_sumr} \\
s^{b}_{nm;a} &= i \sum_{l} ( r_{nl}^{a} s_{lm}^{b} -  s_{nl}^{b} r_{lm}^{a}).
\label{eq:snma_sumr}
\end{align}
The original derivation of Eqns.~\ref{eq:v_sumr}-\ref{eq:rnma_sumr} are presented in Ref.~\onlinecite{Aversa1995}.  Eqns. (\ref{eq:v_sumr})-(\ref{eq:rnma_rnmb_sumr}) and Eqn.~\ref{eq:snma_sumr} can be derived simply by taking matrix elements of 
\begin{align}
[r^a,\hat{H}]/i\hbar &= \hat{v}^{a}, 
\label{eq:comm_rel1} \\
[r^a,\hat{p}^{b}]/i\hbar &=\delta_{ab}, 
\label{eq:comm_rel2}\\
[r^a,r^{b}]&= 0.
\label{eq:comm_rel3} \\
[r^a,s^{b}]&= 0.
\label{eq:comm_rel4}
\end{align}
It is reassuring that the well-known double momentum derivative of energies Eq.(\ref{eq:omega_nab}) is recovered in this approach.\cite{Ashcroft1976} We can also take (covariant) derivatives of any of these to form new ones. For example, we could obtain Eq.(\ref{eq:omega_nab}) and Eq.(\ref{eq:rnma_sumr}) by substituting Eq.(\ref{eq:v_sumr}) into Eq.(\ref{eq:vnm_sumr}) eliminating the velocity, and carefully separating diagonal from off-diagonal terms. Alternatively, Eq.(\ref{eq:omega_nab}) is just the $n=m$ special case of Eq.(\ref{eq:vnm_sumr}). Note that Eq.(\ref{eq:rnma_rnmb_sumr}) and Eq.(\ref{eq:rnma_sumr}) are valid only when the right hand side is evaluated for $n\neq m$. More properly, there should be a factor of $(1-\delta_{nm})$ multiplying the right hand side of these expressions. The covariant derivative of the spin matrix elements Eqn.~\ref{eq:snma_sumr} enters into the BPSE responses for insulators, e.g., Eqn.~\ref{eq:sigma2} and  \ref{eq:eta2}. Note that we implicitly assumed differentiable Bloch wave functions\cite{Panati2007} and the periodic gauge $\pmb{\psi}_{n}(\v{k}+\v{G},\v{r}) = \pmb{\psi}_{n}(\v{k},\v{r})$.

Finally, we comment that taking two momentum derivatives of the gauge-dependent $\langle u_n|\hat{H}_{\v{k}} | u_{m} \rangle= \delta_{nm} E_n$, where $\hat{H}_{\v{k}}= e^{-i\v{k}\cdot\v{r}} \hat{H} e^{i\v{k}\cdot\v{r}}$, gives a similar expression to Eq.(\ref{eq:rnma_sumr}) but with additional term. Caution should be exercised when taking derivatives of matrix elements that depend on the phase of the Bloch wave functions and when using approximate tight-binding Hamiltonians. In our case \ref{eq:v_sumr}-\ref{eq:rnma_sumr} are fixed by the commutation relations Eqs.(\ref{eq:comm_rel1})-(\ref{eq:comm_rel3}) and, in particular, do not depend on the form of the Hamiltonian. 

\section{Symmetry constrains on Bloch matrix elements}
\label{sec:symmetry}
The time-reversed state $\hat{T}|n,i_n,\v{k}\rangle = |n,\bar{i}_n,-\v{k}\rangle \equiv  |\bar{n},-\v{k}\rangle$ is also an eigenstate of the Hamiltonian (Eqn.~\ref{eq:hamiltonian}) if there is time reversal symmetry (TRS), i.e., if $\hat{T}H_0(\v{r},\v{p}) \hat{T}^{-1} = H_0(\v{r},-\v{p}) =H_0(\v{r},\v{p})$.  Here  $\bar{i}_n$ denotes  the spin-flipped spinor and $\hat{T}=-i\sigma^{y}K$ is the time-reversal operator. In this case one can show 
\begin{align}
\v{s}_{nm}(-\v{k}) &= - \v{s}_{\bar{m}\bar{n}}(\v{k}), \\
\v{v}_{nm}(-\v{k}) &= - \v{v}_{\bar{m}\bar{n}}(\v{k}), \\
\v{r}_{nm}(-\v{k}) &= + \v{r}_{\bar{m}\bar{n}}(\v{k}).
\end{align}
The spatially-inverted state $\hat{I}\langle \v{r}|n,i_n,\v{k}\rangle =\langle -\v{r} |n,i_n,\v{k}\rangle = \langle \v{r} |n,i_n,-\v{k}\rangle $ is also an eigenstate of Hamiltonian if there is spatial inversion symmetry (IS), i.e., if $\hat{I}H_0(\v{r},\v{p}) \hat{I}^{-1} =H_0(-\v{r},-\v{p}) =H_0(\v{r},\v{p})$. In this case one can show 
\begin{align}
\v{s}_{nm}(-\v{k}) &= + \v{s}_{nm}(\v{k}) \\
\v{v}_{nm}(-\v{k}) &= - \v{v}_{nm}(\v{k}) \\
\v{r}_{nm}(-\v{k}) &= - \v{r}_{nm}(\v{k}).
\end{align}

\section{FS contributions to linear order conductivity}
\label{sec:Drude_sigma_linear}
We now show that the FS contribution of Eq.(\ref{eq:rho1st22}) recovers the semiclassical Drude conductivity of metals in the relaxation time approximation. This suggest that the spin dynamics arising from the FS is essentially semiclassical. The current to linear order is 
\begin{align}
\v{J}^{(1)} = e\textrm{tr}[\rho^{(1)}\hat{\v{v}}] = \v{J}^{(1i)} + \v{J}^{(1e)}.
\end{align}
Using Eq.(\ref{eq:rho1st22}) and (\ref{eq:rho1st21}) we obtain 

\begin{align}
J^{a(1i)} &= \frac{e^2}{\hbar} \frac{1}{V}\sum_{n\v{k}} \sum_{b\beta} \frac{v_{n}^{a} 
 f_{n;b} }{i\bar{\omega}_{\beta}}  E^{b}_{\beta} e^{-i\omega_{\beta}t} ,
\label{eq:J1i} \\ 
J^{a(1e)} &= \frac{e^2}{\hbar} \frac{1}{V}\sum_{nm\v{k}} \sum_{b\beta}\frac{v_{nm}^{a} r_{mn}^{b} f_{nm}}{\omega_{mn}-\bar{\omega}_{\beta}} E^{b}_{\beta} e^{-i\omega_{\beta}t}.
\label{eq:J1e}
\end{align}
For monocromatic light, the conductivity, defined by
\begin{align}
J^{(1)a}= \sum_b \sigma^{(1)ab}(\omega) E^{b}(\omega) e^{-i \omega t} + c.c.,
\end{align}
has two contributions
\begin{align}
\sigma^{(1)} = \sigma^{(1i)} + \sigma^{(1e)},
\end{align}
where 
\begin{align}
\sigma^{(1i)ab} &= \frac{e^2}{\hbar} \frac{1}{i\bar{\omega}} \frac{1}{V}\sum_{n\v{k}} v^{a}_n f_{n;b},
\label{eq:sigma1i} \\
\sigma^{(1e)ab} &= \frac{e^2}{\hbar} \frac{1}{V}\sum_{nm\v{k}} \frac{v^{a}_{nm} r^{b}_{mn} f_{nm}}{\omega_{mn}-\bar{\omega}}.
\label{eq:sigma1e} 
\end{align}
Eq.(\ref{eq:sigma1i}) can be written as\cite{Ashcroft1976}

\begin{align}
\sigma^{(1i)ab} &=  \frac{e^2}{V}\sum_{n\v{k}} \frac{v^{a}_n v^{b}_n}{1/\tau - i\omega} \left(-\frac{\partial f_{n}}{\partial \mathcal{E}_n} \right),
\end{align}
and is recognized as the standard Drude linear conductivity with a momentum-independent relaxation time. The full quantum mechanical result is usually presented as \cite{Ashcroft1976} 
\begin{align}
\sigma^{(1)ab} &=  - \frac{e^2}{\hbar} \frac{1}{i\bar{\omega}} \frac{1}{V} \sum_{n\v{k}} f_n \bigg[\frac{\hbar}{m}\delta_{ab} \nn\\
&~~~~~~~~~~ - \sum_{l\neq n} (\frac{v_{nl}^a v_{ln}^b}{\omega_{ln}-\bar{\omega}} + \frac{v_{ln}^a v_{nl}^b}{\omega_{ln}+\bar{\omega}} )\bigg].
\label{eq:sigma1am}
\end{align}
To recover this result, note that Eq.(\ref{eq:sigma1i}) can also be written as 
\begin{align}
\sigma^{(1i)ab} &=  - \frac{e^2}{\hbar} \frac{1}{i\bar{\omega}} \frac{1}{V} \sum_{n\v{k}} \omega_{n;ab} f_n,
\end{align}
and using Eq.(\ref{eq:omega_nab}) and Eq.(\ref{eq:sigma1e}) we obtain, after some algebra, Eq.(\ref{eq:sigma1am}). Dissipation in the relaxation time approximation enters by broadening all frequencies in the denominators. In the small frequency limit, interband transitions become the anomalous velocity contribution. Hence, including only the FS term may miss Berry curvature effects.

\section{Interband contribution in the dc-limit: Hall conductivity}
\label{sec:interbad_sigma1e}
The interband term Eq.(\ref{eq:J1e}) contributes to both the dc limit and the finite frequency limit. In the dc limit Eq.(\ref{eq:J1e}) gives the familiar Hall conductivity. This term can be obtained in many ways, but the above formalism allows us to see in a more transparent way various limiting cases. For example, using the identities Eq.(\ref{eq:v_sumr}) and Eq.(\ref{eq:v_Omeganba}), and considering the dissipationless limit $\tau\to \infty$, and the dc limit $\omega\to 0$, we obtain 
\begin{align}
\v{J}^{(1e)}_{dc}= -\frac{e^2}{\hbar} \frac{1}{V}\sum_{n\v{k}} f_n\v{E}_0\times\pmb{\Omega}_n,
\end{align}
which is the the standard Hall current. This result could have been obtained more easily from the intraband current, e.g., Eq.(56) of Ref.~\onlinecite{Fregoso2019}, which includes all contributions to the dc conductivity, including Berry phases, FS contributions, (e.g., $\propto f_{nm;c}$). to any order in a homogeneous electric field.

\section{Second order Drude conductivity}
\label{sec:drude_2nd}
In the relaxation time approximation, the semiclassical distribution in Eq.(\ref{eq:Boltzmann_semicl}) can be written in close form, e.g., Eq.(13.19) of Ref.~\onlinecite{Ashcroft1976}. To obtain the second-order Drude conductivity, simply parametrize the momentum by $\v{k}(t')= \v{k} + e\v{E}_0(t-t')/\hbar$, expand velocity $v^{b}_{n}(\v{k}(t'))$ to first power of $\v{E}_0$, and perform the integrations to obtain the second order semiclassical distribution function 
\begin{align}
g^{(2)}_n = -\frac{e^2 \tau^2}{\hbar} \sum_{bc} \left(-\frac{\partial f_n}{\partial \mathcal{E}_n}\right) v^{b}_{n;c} E^{c}_0 E^{b}_0.
\end{align}
Now substitute into the current 
\begin{align}
j^{(2)a}_{dc} &= \frac{e}{V} \sum_{n\v{k}} v^{a}_n g^{(2)}_n = \sum_{bc} \sigma^{(2)abc}_{dc} E^{c}_0 E^{b}_0
\end{align}
where the conductivity is 
\begin{align}
\sigma^{(2)abc}_{dc} = \frac{e^3 \tau^2}{\hbar^2 V} \sum_{n\v{k}} f_{n;b} v^{a}_{n;c} 
\end{align}
or after integration by parts gives Eq.(\ref{eq:sigma1_2FS}).


%

\end{document}